\documentclass[final]{aipproc}
\layoutstyle{6x9}

\usepackage{amssymb}
\usepackage{pifont}

\def\als{\alpha_{\rm s}}

\begin{document}
\title{Octet contributions in radiative $\Upsilon$ decays}
\author{Xavier Garcia i Tormo}{address={Departament d'Estructura i Constituents de la Mat\`eria. Universitat de Barcelona. Diagonal 647. E-08028 Barcelona. Catalonia. Spain}}

\begin{abstract}
We analyze the end-point region of the photon spectrum in semi-inclusive radiative decays of very heavy quarkonium ($m\als^2 \gg \Lambda_{QCD}$). The $S$- and $P$-wave octet shape functions are calculated. When they are included in the analysis of the photon spectrum of the $\Upsilon (1S)$ system the agreement with data becomes excellent.
\end{abstract}

\begin{flushright}
\small{\tt{UB-ECM-PF 04/28}}
\end{flushright}

\maketitle

The standard Non-Relativistic QCD (NRQCD) factorization \cite{Bodwin:1994jh} (operator product expansion) breaks down at the end-point region of the photon spectrum in semi-inclusive radiative decays of heavy quarkonium; this is due to the fact that collinear degrees of freedom, that are relevant in this kinematic situation, are not included in NRQCD. A proper effective field theory treatment of the end-point region of the spectrum requires, thus, the combination of NRQCD with the Soft-Collinear Effective Theory (SCET) \cite{Bauer:2000yr}. With this approach factorization formulas has been derived for this process and the resummation of Sudakov logarithms has been performed \cite{Fleming:2002sr,Bauer:2001rh}.

\begin{figure}
\centering
\includegraphics{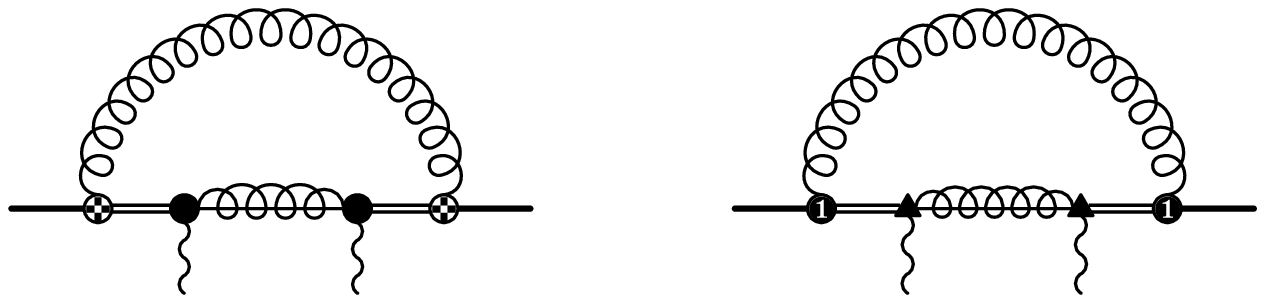}
\caption{\label{dos}Color octet contributions. $\bullet$ represents the color octet S-wave current,
$\blacktriangle$ represents the color octet P-wave current. The notation for the other vertices is
\ding{60}:= ${ig c_F \over \sqrt{N_c T_F}} { \left ( \mathbf{\sigma}_1 - \mathbf{\sigma}_2 \right ) \over 2m } \, {\rm Tr} \left [ T^b {\bf B} \right ]$ and \ding{182}:= ${ig\over \sqrt{N_c T_F}} {\bf x} \, {\rm Tr} \left [ T^b {\bf E} \right ]$. The solid line represents the singlet field, the double line represents the octet field and the gluon with a line inside represents a collinear gluon.
}
\end{figure}

If one considers the initial heavy quarkonium state as a Coulombic state, the octet shape functions (that appear in the factorization formulas) can be calculated with a combination of potential NRQCD (pNRQCD) \cite{Pineda:1997bj} and SCET. The relevant contributions that need to be calculated are depicted in figure \ref{dos}. From the calculation of the two diagrams in that figure we obtain the $S$-wave and $P$-wave octet shape functions in the weak coupling regime, the result is the following:

\begin{equation}
S_{S}(l_+):={4\als (\mu_u)\over 3 \pi N_c} \left({ c_F\over 2m}\right)^2
\int_0^{\infty} dx \left( 2 \psi_{10}( {\bf 0})I_{S}({l_+\over 2} +x)- I_{S}^2({l_+\over 2} +x) \right)
\end{equation}
\begin{displaymath}
S_{P1}(l_+):=
{\als (\mu_u)\over 6 \pi N_c}
\int_0^{\infty}\!\!\!dx\left( 2\psi_{10}( {\bf 0})I_P(\frac{l_+}{2}+x)-I_P^2(\frac{l_+}{2}+x) \right)
\end{displaymath}
\begin{equation}
S_{P2}(l_+):=
{\als (\mu_u)\over 6 \pi N_c}
\int_0^{\infty}\!\!\!dx \frac{8l_+x}{\left(l_++2x\right)^2}\left(
\psi^2_{10}( {\bf 0})-2\psi_{10}( {\bf 0})I_P(\frac{l_+}{2}+x)+I_P^2(\frac{l_+}{2}+x)\right)
\end{equation}
\begin{equation}
I_{S}({k_+\over 2} +x):=m\sqrt{\gamma\over \pi}{\als N_c \over 2}{1\over 1-z'}\left( 1-{2z'\over 1+z'} \;\phantom{}_2F_1\left(-\frac{\lambda}{z'},1,1-\frac{\lambda}{z'},\frac{1-z'}{1+z'}\right)\right)
\end{equation}
\begin{displaymath}
I_{P}({k_+\over 2} +x):=\sqrt{\frac{\gamma^3}{\pi}}
{8\over 3}\left(
2-\lambda \right)\!\!\frac{1}{4(1+z')^3}\Bigg( 2(1+z')(2+z')+(5+3z')(-1+\lambda)+
\end{displaymath}
\begin{displaymath}
+2(-1+\lambda)^2+\frac{1}{(1-z')^2}\left(4z'(1+z')(z'^2-\lambda^2)\left(\!\!-1+\frac{\lambda(1-z')}{(1+z')(z'-\lambda)}+\right.\right.
\end{displaymath}
\begin{equation}
\left.\left.\left.+\phantom{}_2F_1\left(-\frac{\lambda}{z'},1,1-\frac{\lambda}{z'},\frac{1-z'}{1+z'}\right)\right)\right)\right)
\end{equation}
where
\begin{equation}
\gamma=\frac{mC_f\als}{2}\quad z'=\frac{\kappa}{\gamma}\quad-\frac{\kappa^2}{m}=E_1-\frac{k_+}{2}-x\quad\lambda=-\frac{1}{2N_cC_f}\quad E_1=-\frac{\gamma^2}{m}
\end{equation}

These shape function are ultraviolet divergent and need to be regularized and renormalized (see \cite{GarciaiTormo:2004jw} for the details of the procedure); some finite parts coming from a linear divergence must also be subtracted to have a smooth connection with the leading order NRQCD results. %Once this has been done the results can be compared with the experimental data.

Since there is good evidence that the $\Upsilon (1S)$ can be understood as a weak coupling bound state, we compare the results with the experimental data for the $\Upsilon\to X\gamma$ decay \cite{Nemati:1996xy}. The calculation in pNRQCD+SCET is reliable in the region $z\in\left[0.7,0.95\right]$ (where $z=2E_{\gamma}/M$, $M$ being the mass of the heavy quarkonium state), in order to have a reliable prediction for the whole spectrum we adapt the interpolation formula used in \cite{Fleming:2002sr} for the inclusion of the color octet contributions, that is
\begin{equation}\label{int}
\frac{1}{\Gamma_0}\frac{d\Gamma_{int}}{dz}\!\!=\!\!\frac{1}{\Gamma_0}\frac{d\Gamma_{LO}^{dir}}{dz}+\!\!\left(\!\!\frac{1}{\Gamma_0}\frac{d\Gamma_{resum}^{sing}}{dz}-z\!\!\right)\!\!+\!\!\left(\!\!\frac{1}{\Gamma_0}\frac{d\Gamma_{resum}^{oct}}{dz}-z\left(4+2\log(1-z)\right)(1-z)\!\!\right)
\end{equation}
Furthermore the low photon energy region of the spectrum $z\lesssim 0.4$ is dominated by the so called fragmentation contributions; these contributions must be added to the previous ones to obtain the total decay rate; we will use the formulas presented in section II.B of the first reference in \cite{Fleming:2002sr} and the evaluation of the NRQCD color octet matrix elements derived in the appendix of \cite{GarciaiTormo:2004jw} to obtain these fragmentation contributions. %(a detailed discussion of these last two paragraphs will be given in \cite{aviat}).

\begin{figure}
\centering
\includegraphics{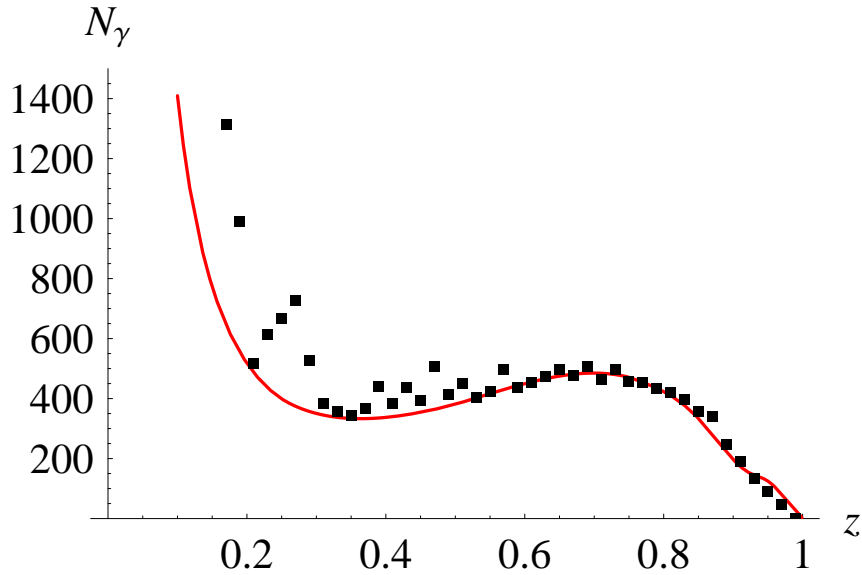}
\caption{\label{grllarg}Photon spectrum in semi-inclusive $\Upsilon$ decay. The points are the CLEO data \cite{Nemati:1996xy}, the solid line is the theoretical prediction described in the text.
}
\end{figure}

The comparison with data is depicted in figure \ref{grllarg}, the solid curve consists of the color octet contribution (with the Sudakov resummation performed in \cite{Bauer:2001rh} included) and the color singlet contribution \cite{Fleming:2002sr}, interpolated according to \eqref{int} with the leading order NRQCD result (which is reliable away from the end-point), plus the fragmentation contributions. No error bars are presented in this plot (neither for the experimental data points nor for the theoretical curve) since the aim is just to illustrate that a good description of the spectrum is possible (a detailed description of this analysis will be presented in \cite{aviat}). Taking into account that the data points below $z=0.4$ have large errors we can affirm that the agreement of the theoretical curve with data is excellent.

\begin{theacknowledgments}
It is a pleasure to thank J. Soto for past and present collaboration on this subject. Financial support from a CICYT-INFN 2003 collaboration contract,
the MCyT and Feder (Spain) grant FPA2001-3598, the CIRIT (Catalonia) grant 2001SGR-00065, the Departament d'Universitats, Recerca i Societat de la Informaci\'o of the Generalitat de Catalunya and the Acciones Integradas Espa\~na-Italia, project HI2003-0362 is acknowledged.
\end{theacknowledgments}

\end{document}